\documentclass[10pt]{article}
\usepackage{amsfonts}
\usepackage{amssymb}

\newcommand{\1}{{\sf 1}}

\renewcommand{\H}{\bold H}
\newcommand{\K}{{\bold K}}
\renewcommand{\L}{\bold L}
\newcommand{\M}{\bold M}

\newcommand{\X}{\bold X}

\newcommand{\tr}{\, \mbox{Tr}\, }
\newcommand{\Tr}{\, \mbox{Tr}\,}

\newcommand{\spn}{\, \mbox{span}\,}

\newcounter{thaler}
\newenvironment{mlist}{\begin{list}{\arabic{thaler}}%
{\usecounter{thaler}
\setlength{\rightmargin}{\leftmargin}
\topsep=0pt
\itemsep=0pt
\parskip=0pt
\parsep=0pt
}}{\end{list}}

\newenvironment{boxx}%
{\begin{center}%
\begin{tabular}{|c|} \hline \\ $\displaystyle \ \ \ }%
{\ \ \ $\\ \\ \hline \end{tabular} \end{center}}

\newcounter{thingy}
\renewcommand{\thethingy}{\arabic{section}.\arabic{thingy} }

\renewcommand{\section}[1]{ \refstepcounter{section}
\setcounter{thingy}{0}
\underline{ \bf \arabic{section}~#1 }}{} 

\newcommand{\textitem}[1]{ \refstepcounter{thingy}
 {\bf \thethingy #1.}}{} 

\newenvironment{examples}
{\refstepcounter{thingy} {\bf \thethingy Examples.} }{}

\newenvironment{example}
{ \refstepcounter{thingy}{\bf \thethingy Example. } }{}

\newenvironment{definition}
{ \refstepcounter{thingy}{\bf \thethingy Definition} }{}

\newenvironment{corollary}
{ \refstepcounter{thingy}{\bf \thethingy Corollary} }{}

\newenvironment{theorem}
{ \refstepcounter{thingy}{\bf \thethingy Theorem} \em }{}

\newenvironment{proposition}
{ \refstepcounter{thingy}{\bf \thethingy Proposition} \em }{}

\newenvironment{lemma}
{ \refstepcounter{thingy}{\bf \thethingy Lemma} \em }{}

\newcommand{\ket}[1]{| #1 \rangle}
\newcommand{\bra}[1]{\langle #1 |}
\newcommand{\proj}[1]{\ket{#1}\! \bra{#1}}
\newcommand{\outerp}[2]{\ket{#1}\! \bra{#2}}

\newcommand{\inner}[2]{\langle #1, #2 \rangle}

\newcommand{\text}{\mbox}
\renewcommand{\Tr}{\tr}

\newcommand{\tp}[1]{{#1}^{t}}

\newcommand{\Ch}{\text{Ch}}
\newcommand{\w}{{\bf w}}
\newcommand{\forward}[2]{\stackrel{\longrightarrow}{\frak{#1} \frak{#2}}}
\newcommand{\backward}[2]{\stackrel{\longleftarrow}{\frak{#1} \frak{#2}}}
\newcommand{\leftright}[2]{\stackrel{\longleftrightarrow}{\frak{#1} \frak{#2}}}

\newcommand{\J}{{\bf J}}

\def\openone{\leavevmode\hbox{\small1\kern-3.8pt\normalsize1}}
\def\RR{{\rm I\kern-.2emR}}
\def\tr{{\rm Tr}\; }
\def\ce{{\cal E}}

\def\fb{{\frak B}}

\def\fa{{\frak A}}

\def\idop{{\rm~id~}}

\newcommand{\Id}{\text{id}}

\renewcommand{\inner}[2]{ \langle #1 | #2 \rangle}
\newcommand{\dmelement}[2]{ \langle #1 | #2 | #1 \rangle}
\newcommand{\matelement}[3]{ \langle #1 | #2 | #3 \rangle}
\newcommand{\beq}{\begin{equation}}
\newcommand{\eeq}{\end{equation}}
\newcommand{\beqa}{\begin{eqnarray}}
\newcommand{\eeqa}{\end{eqnarray}}
\newcommand{\ora}[1]{\stackrel{ \longrightarrow }{ #1 }}

\def\QED{\mbox{\rule[0pt]{1.5ex}{1.5ex}}}

\begin{document}

\begin{center}{\large \bf Influence-free states on compound quantum systems} 

Howard Barnum\footnote{CCS-3: Modelling, Algorithms, and Informatics.  Mail 
Stop B256, Los Alamos National Laboratory, Los Alamos, NM 87545; 
{\tt barnum@lanl.gov}}, 
Christopher A. Fuchs\footnote{Quantum Information and Optics Research, Bell
Labs, Lucent Technologies, 600-700 Mountain Avenue, Murray Hill, New 
Jersey 07975 USA; {\tt cafuchs@research.bell-labs.com}}, 
Joseph M. Renes\footnote{Instit\"ut f\"ur Theoretische
Physik I, Universit\"at Erlangen-N\"urnberg Staudtstrasse 7/B2, 
91058 Erlangen, Germany; {\tt jrenes@optik.uni-erlangen.de}} 
and Alexander Wilce\footnote{Department of Mathematical Sciences,
University of Susquehanna, Selinsgrove, PA 17870 USA; {\tt wilce@susqu.edu}}.
 
{\sl Version of July 11, 2005}\end{center} 

\begin{abstract} 
Consider two spatially separated agents, Alice and Bob,
each of whom is able to make
local quantum measurements, and who can communicate with each other
over a purely classical channel. As has been pointed out by a number
of authors, the set of mathematically consistent joint probability
assignments (``states'') for such a system is properly
larger than the set of quantum-mechanical mixed states for the joint
Alice-Bob system.  Indeed, it is canonically isomorphic to the set of
positive (but not necessarily completely positive) linear maps
${\cal L}(\H_A) \rightarrow {\cal L}(\H_B)$  from the bounded linear
operators on Alice's
Hilbert space to those on Bob's, satisfying $\tr(\phi(\1)) =
1$.  The present paper explores the properties of these
states.  We review what is known, including the fact that
allowing classical communication between parties is equivalent to
enforcing ``no-instantaneous-signalling'' (``no--influence'')
in the direction opposite the communication.  We establish
that in the subclass of  ``decomposable'' states,
i.e. convex combinations of positive states with ones whose partial
transpose is positive, the extremal (i.e.  pure)
states are just the extremal positive and extremal
partial-transpose-positive states.  And we show that two such states,
shared by the same pair of parties, cannot necessarily 
be combined as independent
states (their tensor product) if the full set of quantum operations is
allowed locally to each party: this sort of 
coupling does not mix well with the standard quantum one.  We use
a framework of ``test spaces'' and states on these, well suited
for exhibiting the analogies and deviations of empirical probabilistic
theories (such as quantum mechanics) 
from classical probability theory.  This leads to a deeper understanding
of some analogies between quantum mechanics and Bayesian probability theory.
For example, the existence of a ``most Bayesian'' quantum rule for
updating states after measurement, and its association with the 
situation when information on one system is gained by measuring
another, is a case of a general proposition holding for test spaces
combined subject to the no-signalling requirement.
\end{abstract}

\newpage
\underline{\bf Outline}
\begin{mlist}
\item[(1)] Introduction
\item[(2)] States and weights on test spaces
\item[(3)] States on coupled systems; influence-freedom
\item[(4)] Operator representations of influence-free states
\item[(5)] Decomposable states 
\item[(7)] Teleportation

\end{mlist}
       
\section{Introduction}

In the simplest formulation of orthodox non-relativistic quantum mechanics, a 
physical system is represented by a separable, complex Hilbert space $\H$. The 
possible outcomes of (maximal) discrete measurements on such a system are 
represented by unit vectors in $\H$, with each orthonormal basis representing 
the set of mutually exclusive possible outcomes of a given such measurement. 
By Gleason's theorem \cite{Gleason}, in dimension greater than two 
any consistent assignment of probabilities to all of 
these outcomes -- i.e., any assignment of values in $[0,1]$ to each unit 
vector that sums to $1$ over each orthonormal basis -- arises uniquely from a 
density operator, i.e., a trace-1, positive self-adjoint operator $W$, by the 
formula $x \mapsto \dmelement{x}{W}$ ($x$ a unit vector in $\H$). 

A slightly more general formulation associates measurement outcomes to
``effects,'' operators belonging to the unit interval (between $0$ and
the identity operator $I$) in the positive semidefinite ordering, and
mutually exclusive possible outcomes of a given measurement to
discrete resolutions of unity into effects $e_i$: $\sum_i e_i = I$.
An analogue \cite{Busch03a,CFRM03a} of Gleason's theorem (cf. also
the account in \cite{Fuchs2001a, Fuchs2002a}), easier 
because its essence
is the self-duality of the cone of positive semidefinite operators,
tells us that consistent probability assignments still correspond to
density matrices $W$, with probabilities given by $e \mapsto \tr e W$.
With a little more mathematical sophistication, one can generalize
this a little further to allow certain measurements having a continuum
of outcomes: this is the theory of POVM's (positive operator valued
measures) and the associated generalized observables.  As is common
in quantum information theory, we will also refer to the discrete
resolutions of unity introduced above as POVMs, and will mostly avoid
the continuous case in what follows.

A number of recent papers in quantum information theory (notably ones
by C.~A. Fuchs \cite{Fuchs2001a, Fuchs2002a} and N. Wallach \cite{Wallach})
have considered situations in which 
measurements are made on 
a pair of systems, and proved ``unentangled Gleason theorems'' about 
the representation of probabilities in such situations. 
The joint system is represented by the tensor product $\H = \H_{1}
\otimes \H_{2}$. 
But the measurements are restricted to have all of their 
outcomes correspond to effects that are tensor products $e \otimes f$.
(In the simpler situation without effects, the analogue would be to 
restrict measurements to have all outcomes corresponding to tensor
product vectors, $\ket{x} \ket{y}$.)
\cite{Wallach} considers general such measurements, while 
\cite{Fuchs2001a, Fuchs2002a} restrict
further, to 
a pair of
(say, space-like) separated observers who make local quantum measurements,
and share information over a classical channel. 
The question
arises: do all mathematically consistent probability assignments to
pure tensors arise from standard quantum states, i.e., density
operators on $\H$?  The answer is no. In the case that $\H_{1}$ and
$\H_{2}$ are finite dimensional, one can indeed find a trace-1
self-adjoint operator $W$ on $\H$ such that the probability to obtain
outcome $e \otimes f$ (or $\ket{x}\ket{y}$, in the simpler model) 
is given by $\tr (W (e \otimes f))$ (or 
$\bra{y}\bra{x} W \ket{x}\ket{y}$, in the simpler model).  
However, the operator need not be positive. Rather, it
need only be {\em positive on pure tensors} (POPT): 
\[\bra{y}\bra{x} W \ket{x}\ket{y}\ \geq 0 \ \ \text{for all~} x \in
\H_{1}, y \in \H_{2}.\]   (This is equivalent to positivity of 
$\tr (W (e \otimes f))$ for all effects $e,f$.) 
Thus, the theory of such ``local'' probability
assignments outruns ordinary quantum statistics. Put another way:
there exist (mathematically, at any rate) non-quantum mechanical
states on coupled quantum systems having quantum-mechanical marginals.
                                       
In fact, such states were also considered in essentially this same
 context, by Kl\"{a}y, Randall and Foulis \cite{Klay-Randall-Foulis} (KFR)
(see also \cite{Klay, Wilce90, Wilce92}), and also by Rudolf and
 Maitland-Wright \cite{Rudolf-Wright} in connection with decoherence
 functionals.  In this paper, we present much of the available 
information on the structure and properties
 of such states and their representing operators, along with some results
we have not seen elsewhere.
 
In Sections 2 and 3, following Foulis and Randall 
\cite{Foulis-Randall80a, Foulis-Randall80b} 
and others, we introduce a fairly general
framework, that of test spaces, for discussing statistical models such
as the Hilbert-space model of quantum mechanics, or classical
probability theory.  In particular, we review the ``tensor product" of
Foulis and Randall (FR), 
and prove a simple representation theorem for states thereon
\cite{Wilce92}.  In Section 4, we give a proof of the KFR
``unentangled Gleason theorem", showing that states on the FR tensor
product of two frame manuals are representable by positive maps,
or equivalently, by POPT operators (``Choi matrices" of the maps).
While test spaces capture the simple version of quantum mechanics in
terms of ``von Neumann'' measurements whose outcomes correspond to
projectors, they are not quite general enough to encompass POVMs.
Nevertheless the content of the ``unentangled Gleason's theorems''
based on test spaces is essentially the same as that of the versions
based on concrete Hilbert-space effects and ``unentangled POVMs.''
Since the relation between test space and effect-test space versions has
little bearing on the underlying physics and information-processing
content we are concerned to highlight, we will say little more about
it here, and discuss it in an extended version of the paper in which
the relationship between test space based models and related algebraic
structures, and effect-based models and their related algebraic
structures, can be investigated at length.

In Section 5, we discuss decomposable
states, proving that extreme decomposable states are either extreme CP
or extreme co-CP states.  Section 6 develops a variant of the
standard quantum teleportation protocol that shows that POPT states on
subsystems do not generally extend to POPT states on full systems. To
keep the main flow of the discussion clearly in view, some of the
background material has been placed in an appendix. 

\section{States and Weights on Test Spaces}
                                                      
Since we'll be venturing outside of the usual conceptual shell of 
quantum probability theory, it will be helpful to establish, by way of a 
scaffolding, a language for discussing statistical models generally. 
Much of what follows could be framed in terms of abstract convex sets, 
ordered vector spaces, etc.   However, we will use
the language of states on test spaces, which is almost 
equally general but slightly more concrete. 
 
\textitem{Test spaces} A {\em test space} (sometimes called a {\em manual})
is just a collection of classical 
discrete sample spaces, possibly overlapping. To be more formal, let's agree 
that a test space is a pair $(X,{\frak A})$ where $X$ is a set  and ${\frak A}$ is a covering of $X$ by non-empty subsets. Each set $E \in {\frak A}$ is supposed to represent the set of possible {\em outcomes} of some experiment, measurement, operation or 
{\em test}.  Accordingly, a {\em state} (or probability weight) on a test 
space $(X,{\frak A})$ is a map $\omega : X \rightarrow [0,1]$ summing 
independently to $1$ over each test $E \in {\frak A}$. We write the 
space of all states on a test space $(X, \fa)$ as $\Omega(X,\fa)$. 

 
\begin{examples} 
Of course, if there is only one test -- that is,
if our test space has the form $(E,\{E\})$ -- we recover the framework
of discrete classical probability theory. If $X$ is the set of
non-empty measurable subsets of a measurable space $S$ and ${\frak A}$
is the set of countable partitions of $S$ into measurable subsets, we
recover the framework of classical probability theory. To recover the
usual framework of quantum probability theory, let $\H$ be a Hilbert
space, let $X(\H)$ denote $\H$'s unit sphere, and let ${\frak F}(\H)$
denote the set of frames (maximal orthonormal subsets) of $\H$. Then
$(X,{\frak F})$ is a {\em quantum test space} (or {\em frame manual});
Gleason's theorem identifies the states on $(X,{\frak F})$ with
density operators on $\H$. 
\end{examples}

\textitem{The Space of Signed Weights}
It is sometimes useful to consider states that are not normalized,
and can also be useful to consider states as belonging to a
vector space of real-valued functions on $X$ (the one they 
generate, in fact).  So we make the following definitions.
A {\em positive weight} on $(X,{\frak A})$ is a positive multiple of a state,
i.e., a function $f : X \rightarrow {\Bbb R}$ such that (i) $f(x) \geq
0$ for all $x \in X$, and (ii) there is some constant $K \geq 0$ with
$\sum_{x \in E} f(x) = K$ for every $E \in {\frak A}$.  By a {\em
signed weight}, we mean a function of the form $f - g$, where $f$ and
$g$ are positive weights. The set of signed weights on $(X,{\frak
A})$, denoted by $V(X,{\frak A})$, is a subspace of ${\Bbb R}^{X}$ --
in fact, just the span of $\Omega(X,{\frak A})$.\footnote{This has a
natural base-norm, with respect to which it's complete. Since our
interest here is mainly finite-dimensional, we needn't worry too much
about this.}  We write $V^{\ast}(X, {\frak A})$ for the vector space
dual to $V(X,{\frak A})$, i.e. the space of linear functionals 
$V(X,{\frak A}) \mapsto {\bf R}$.  

Note that every outcome $x \in X$ determines a positive linear
evaluation functional $\phi_x \in V^{\ast}(X,{\frak A})$ given by
$\phi_{x}(\omega) = \omega(x)$. Since these obviously separate points
of $V(X,{\frak A})$, we have for $V$ finite dimensional that
$V^{\ast}(X,{\frak A})$ is spanned by the $\phi_x$.

\textitem{Vector-Valued Weights.}
It's also useful to consider more general vector-valued weights on a test space $(X,{\frak A})$. Let $\M$ be a Banach space. Given a function $\omega : X \rightarrow \M$, we define the {\em variation} of $f$ to be 
\[\|\omega \| := \sup_{E \in {\frak A}} \sum_{x \in E} \|\omega(x)\|.\]
(understanding that $\|f\| = \infty$ if the sup does not exist.) We call $f$ 
a $\M$-valued weight iff  (i) $\|f\| < \infty$, and (ii) there exists a 
constant $\w \in \M$ such that for any $E$, 
$\sum_{x \in E} \omega(x)$ converges in norm to 
$\w$.  The space of $\M$-valued weights is a Banach space with 
respect to the variation norm \cite{Wilce93}. \footnote{Of course, we have $V(X) \leq W(X)$. In general, the inclusion is proper: not every real-valued weight is the difference of positive 
weights.}\\ 

\textitem{Effect-test Spaces}  Although we will not delve deeply into
the relation between effect-test spaces and standard test-spaces, and
related objects associated with them, we will introduce some basic concepts
in order to make a few remarks on these relations later on.  

Effect-test spaces, or E-test spaces, are
similar to test spaces, but allow for the possibility that an outcome may
occur in a test {\em with multiplicity}.  For example, in 
the general formulation of
quantum mechanics in terms of POVMs, mentioned above, this is possible, 
so E-test spaces rather than just test spaces are necessary if one wants
to encompass it.  Confining ourselves for simplicity to {\em locally discrete}
E-test spaces, (so in the quantum case we can encompass discrete, 
but not yet continuous POVMs) we may define
an E-test space $(Z,{\frak D})$ 
as a set ${\frak D}$ of (possibly overlapping) multisets of elements
of $Z$.  For our purposes
a multiset $s$ 
is just a map from the ground set $Z$ to the nonnegative integers.  The ground
set elements $x$ on which the map takes a nonzero value $s(x)$
are thought of as occuring in the multiset $s(x)$ times.  Thus an ordinary
set (with elements drawn from the ground set) is just a multiset for which 
$s$ is $0,1$-valued.  (In this case $s$ is what is called the 
characteristic function of the set.)  States are maps $\omega: Z \rightarrow [0,1]$ such
that for every $s \in {\frak A}$, $\sum_{x \in Z} s(x) \omega(x) = 1$.

\section{States on Coupled Systems; Influence-Freedom} 

In what follows, we consider an arbitrary pair 
of test spaces $(X,{\frak A})$ and $(Y, {\frak B})$. We'll think of these as 
associated with two ``agents" (Alice and Bob, say) located at different sites, but 
able to communicate with one another. 

We'll use a juxtapositive notation for ordered pairs of 
outcomes, writing $xy$ for $(x,y)$, and, for sets $A \subseteq X$ and $B 
\subseteq Y$, $AB$ instead of $A \times B$ for 
$\{ xy | x \in A \ \text{and} \ y \in B\}$.
Thus, the outcome set for the join test in which Alice performs test $E \in {\frak A}$ and Bob performs test $F \in {\frak B}$ is $EF$. The set of all such ``product tests" defines a new test space:
         
\begin{definition} The {\em Cartesian product} of $(X,{\frak A})$ and $(Y,{\frak B})$ is the test space $(XY, {\frak  A} \times {\frak B})$ where ${\frak A} \times {\frak B} = \{ EF | E \in {\frak A} \ \text{and} \ F \in {\frak B}.\}$.  
                            
If $\omega$ is a state on $(XY,{\frak A} \times {\frak B})$  
and $E \in {\frak A}$, we 
can define the {\em marginal state} ${}^{E}\omega_{\frak B}$ on $(Y,{\frak B}$ 
by 
\[{}^E\omega_\fb(y) := \omega(Ey) \equiv \sum_{x \in E} \omega(x,y).\]
The marginal state ${}_\fa \omega^F$, $x \in X$ and $F \in {\frak B}$, 
is defined similarly.
\end{definition}

\begin{definition}
Call a state $\omega$ on $(X \times Y, {\frak A} \times 
{\frak B})$ {\em influence-free} iff the marginal states 
${}^E\omega_\fb : y \mapsto \omega(Ey)$ and ${}_\fa {\omega}^F : x 
\mapsto \omega(xF)$  are independent of $E$ and of $F$, respectively. In this 
case, we write $_{\frak A} \omega$ and $\omega_{\frak B}$ for these 
marginals. We'll write  $\Omega_{\rm free}(X,Y)$ for the set of influence-free states on 
$X \times Y$. 
\end{definition}

``Influence-free'' here should emphatically {\em not} be thought to
imply what quantum information theorists sometimes call separable,
i.e. unentangled; entangled quantum states are still influence-free.
Rather, influence-free states are those that do not permit the slightest
fraction of a bit of instantaneous signalling: one party cannot
influence the other party's probabilities merely by his or her choice
of measurement.  Influence-freedom is a conjunction of two conditions,
which we'll call $\leftarrow$-influence-freedom and
$\rightarrow$-influence-freedom: no influence of system 1 on system 2,
and vice versa.
           
\textitem{Classical Communication} Suppose that the two systems
represented by $(X,{\frak A})$ and $(Y,{\frak B})$ belong to two
agents, Alice and Bob, who are located at spatially remote sites but
can communicate with one another via a classical channel. Then joint
experiments more complex than $EF$ are possible. For instance, Alice
may perform test $E$, and, upon securing outcome $x \in E$, instruct
Bob to perform a test $F_{x} \in {\frak B}$. This yields a compound
test with outcome set $\bigcup_{x \in E} xF_{x}$.

Let $\stackrel{\longrightarrow}{{\frak A}{\frak B}}$ denote the set of all such 
compound experiments, and let $\stackrel{\longleftarrow} {{\frak A}{\frak 
B}}$ denote the corresponding set of two-stage tests $\bigcup_{y \in F} E_{y} 
y$ in which Bob performs the initial test.

\begin{definition}\label{def: influence-free product}
\beqa
{\frak A} {\frak B} := \forward{A}{B} \cup \backward{A}{B},
\eeqa
\end{definition}

In other words, ${\frak A}{\frak B}$ is the set of two-stage
experiments initiated at either Alice's or Bob's site. Note that
$\bigcup{\frak A}{\frak B} = XY$, and also that ${\frak A} \times
{\frak B} \subseteq {\frak A} {\frak B}$.\footnote{Indeed, ${\frak A}
\times {\frak B} = \stackrel{\longrightarrow}{{\frak A}{\frak B}} \cap
\stackrel{\longleftarrow}{{\frak A}{\frak B}}$).} Thus, the
state-space of $(XY,{\frak A}{\frak B})$ is a convex subset of the
state-space of $(XY,{\frak A} \times {\frak B})$.   The following was
shown in \cite{Foulis-Randall80b, Randall-Foulis, Klay}), and we
will review the simple proof of a proposition that implies it.
(Note that in \cite{Foulis-Randall80a, Foulis-Randall80b, Klay} 
the notation $\fa \fb$ is used for what we call $\fa \times \fb$, while 
$\leftright{A}{B}$ is used for what we call $\fa \fb$;  also the term
``interference'' is sometimes interchangeably with ``influence'' as
we have defined it.)

\begin{corollary} Let $\omega$ be a state on 
$(XY,{\frak A} \times {\frak B})$. Then the following are equivalent: 
\begin{mlist}
\item[(a)] $\omega$ is a state on $(XY,{\frak A}{\frak B})$; 
\item[(b)] $\omega$ is influence-free.
\end{mlist}
\end{corollary}

This is an immediate corollary of the following easy theorem, which can be
interpreted as saying that allowing classical communication in one
direction enforces that the states be influence-free in the opposite
direction.

\begin{theorem}(Randall and Foulis \cite{Randall-Foulis}, Lemma 2.8) 
\label{theorem: FR}
Let $\omega$ be a state on 
$(XY,{\frak A} \times {\frak B})$. Then the following are equivalent: 
\begin{mlist}
\item[(a)] $\omega$ is a state on $(XY,\stackrel{\longrightarrow}{{\frak A}{\frak B}})$; 
\item[(b)] $\omega$ is $\leftarrow$-influence-free.
\end{mlist}
\end{theorem}

{\bf Proof of Theorem~\ref{theorem: FR} }

\begin{proof}
We first show ``only if.''  
Let $\omega \in \Omega(\ora{{\frak{A}} {\frak{B}}})$.
A test  $T \in \ora{{\frak{A}} {\frak{B}}}$ involves
performing a test $Z \in \fa$ and proceeds,
on obtaining
outcome $z \in Z$, by performing  
$W_z \in \fb$. 
Thus, for any chosen $x \in Z$, 
\beqa
1 = \omega(T) = \sum_{z \in Z-x} 
\sum_{y \in W_z} \omega(z,y)
+  \sum_{y \in W_x} \omega(x,y)\;.
\eeqa
In other words,
\beqa
\sum_{y \in W_x} \omega(x,y) = 
1 - \sum_{z \in {T - x}}  \sum_{y \in W(z)}
\omega(z,y) \;.
\eeqa
The RHS is independent of $W_x$, while the LHS defines 
${}^{W_x}\omega_\fb(x)$.   Since the construction
can be done (varying $Z$ if necessary)
for any $x$ and any choice of test $W$ as $W_x$ 
this establishes that $^{W}\omega_\fa$ is independent of $W$,
i.e. $\omega$ is influence-free.  

To show ``if'':  suppose we have an influence-free state 
$\omega$ on  ${\frak{A}} \times {\frak{B}}$.  We just have to show it 
adds up to one on tests in $\ora{{\frak{A}} {\frak{B}}}$.
For an arbitrary such test $T$ consisting of performing $Z \in \fa$
followed, conditional on result $z \in Z$, by performing
$W_z \in \fb$, we have: 
\beqa
\sum_{(x,y) \in T} & = & \sum_{z \in Z} 
\sum_{y \in W_z} \omega(x,y) \\
& = & \sum_{z \in Z}~~ {}_\fa^{W_z} \omega(x) \\
& = & \sum_{z \in Z} ~~ _\fa \omega(x) = 1
\eeqa
The next-to-last equality is due to $\leftarrow$-influence-freedom
of $\omega$.
\end{proof}

The test space $\frak{A}\frak{B}$ has sometimes been called the 
Foulis-Randall product or bilateral tensor product of $\frak{A}$ and
$\frak{B}$.  We will sometimes call it the ``free no-signalling,''
or ``FNS'' product.  

\textitem{Digression:  Conditional dynamics at a distance;  marginalization
and Bayesian updating}

We may interpret the existence of a marginal state as giving 
a ``conditional dynamics'' of Bob's state, caused 
by Alice measuring and obtaining the outcome she does.  
This ``conditional dynamics''
is nothing but ordinary probabilistic
conditionalization.  That is, letting $\omega^{AB}$ be the initial
Alice-Bob state, $x$ any Alice outcome (such as a Bell-measurement
outcome), we may ascribe to Bob the post-measurement state $\omega_x^B$, 
defined by $\omega_x^B(y) := \omega^{AB}(xy)/ \sum_{y \in T} \omega^{AB}(xy)$.
The RHS refers to a test $T$, but it does not depend on the choice of
this test:  the fact that it does not is precisely $\leftarrow$-no-signalling.
We may therefore write this denominator as $\omega^A(x)$ (we have a 
well-defined marginal state $\omega^A$), and the expression for Bob's
conditional state then looks just like ordinary probabilistic
conditionalization:
\beq
\label{genconditionalization}
\omega^B_x(y)  = \omega^{AB}(xy)/\omega^{A}(x)\;.
\eeq

In \cite{Fuchs2001a, Fuchs2002a} an analogy was made between
decomposing a quantum mechanical density operator $\rho$ as a convex
combination
\beq \label{qbayes}
\sum_b p(b) \rho_b = \rho
\eeq
of density operators, and 
``refining one's knowledge'' of a quantum system 
\cite{Fuchs2001a, Fuchs2002a}.  
Due to the concavity of von Neumann
entropy, one might say (using 
entropy---or indeed any other unitarily invariant extension of
a Schur-concave function, cf. \cite{FJ01a}--as a measure of one's
knowledge of a system) that if one expects to learn a value of 
$b$, and thereupon have knowledge represented by 
$\rho_b$ about the system, then one expects on average to gain 
information about a quantum system.  
In fact, if $\rho$ is taken to represent the state of a quantum 
system before a measurement, for any  additive decomposition of
the form (\ref{qbayes}) there exists a measurement with conditional
dynamics---a
set of completely positive  maps $\ce_b$ summing to a trace-preserving
one $\ce$---such that the post-measurement states $\ce_b(\rho)$
conditional on outcome $b$ of the measurement are $\rho_b$.  Of course,
this same measurement-with-dynamics will not in general have the same
property for {\em other} input density matrices $\sigma$.
Conditional dynamics 
such that the conditional density operators sum to the initial
one are by no means the general form of conditional quantum
dynamics, and (as shown in \cite{FJ01a})
the fact that one gains information (in the above sense)
on average also holds for many other conditional 
dynamics...indeed, for {\em any} ``efficient''  measurement of a 
POVM on a quantum system. 
(``Efficient'' here means that the 
post-measurement states conditional on measurement outcomes are 
pure, so that there is no information extracted in the interaction of
system with apparatus and environment during measurement, that is 
not reflected in the measurement outcome.)
 But (as argued in \cite{Fuchs2001a, Fuchs2002a}) the dynamics
\beqa
\rho \stackrel{b}{\rightarrow} \rho_b \nonumber \\
prob(b) = \tr \rho_b
\eeqa 
meaning that conditional on the measurement outcome
$b$, which occurs with probability $\tr \rho_b$, 
$\rho$ evolves to $\rho_b$,
are especially
close to those obtained classically via Bayes' rule.
(There it is also shown that it is always possible, in 
quantum mechanics, to measure in such a way that these conditional
dynamics occur, although the ``instrument'' (collection of 
completely positive maps, one for each outcome $b$ of the POVM) 
achieving this will depend not only on the POVM but also
on the initial density operator $\rho$.)  
The analogy with Bayesian updating is as follows:  let 
there be random variables $A, B$ taking values in a finite
set, and consider an initial joint distribution $p(A,B)$.
We have an initial distribution $p(A)$, and wish to obtain
information about $A$ by observing $B$.  Then, the conditional
distribution $p_b(A)$ (more frequently written $p(A|b)$
is defined via 
\beq \label{cbayesrule}
p_b(a) = p(ab)/\sum_a p(ab)\;.
\eeq
Usually 
one defines $p(b) := \sum_a p(ab)$ and thus it
holds that 
\beqa \label{cbayes}
\sum_b p(b) p_b(A) = p(A)\;.
\eeqa  
This form of Bayes'
rule is the ``classical'' analogue of
(\ref{qbayes}).  In fact, the analogy can be sharpened:
(\ref{qbayes}) implies that (calling the observable whose
values are the $b$, and which is presumed to be measured
with the conditional dynamics $\rho \stackrel{b}{\rightarrow} 
\rho_b$ with $\rho_b$ satisfying (\ref{qbayes})),
for {\em any} choice of post-measurement observable $Y$, there
exists a joint distribution $p(B,Y)$ determined by $\rho$, 
such that the probabilities for the values $a$ taken by $Y$
conditional on outcome $b$, are given by (\ref{cbayesrule})
and thus satisfy (\ref{cbayes}) (with
$Y$ substituted for $A$ in the latter, of course).  

When do quantum systems exhibit such ``closest-to-Bayesian''
dynamics?  As shown in \cite{Fuchs2001a, Fuchs2002a}, one such situation is when 
information about a quantum 
system $A$ is obtained by measuring another system  $B$ that may
have
been entangled with it.  In that case the
reduced density matrix of $A$ (calculated from the joint
state $\rho^{AB}$) is updated (on learning the result of 
a measurement on $B$) precisely according to a rule of the
form  $\rho \stackrel{b, p(b)}{\rightarrow} \rho_b$ 
satisfying (\ref{qbayes}).  

We may understand this result as a special case of a general
relationship in FNS tensor products.  Although the quantum-mechanical
tensor product is not the FNS tensor product, but rather permits
more measurements and fewer states, the measurements involved in 
the result do belong to those permitted in the FNS tensor 
product, and for these measurements and states, the quantum 
probabilities and the FNS ones coincide.

Looking again at (\ref{genconditionalization}), we see that
we may rewrite the effect of conditionalizing on a $B$ measurement
result as:
\beqa \label{gbayesrule}
\omega^B \stackrel{a}{\rightarrow}  \omega_a^B \nonumber \\
prob(a) = {\omega^A(a)}
\eeqa
where 
\beqa \label{gbayes}
\sum_a \omega^A(a) \omega_a^B = \omega^B\;.
\eeqa
These are general analogues of (\ref{cbayesrule}) and (\ref{cbayes})
(and (\ref{gbayes}) is a general analogue of (\ref{qbayes})).  
They can be summarized (and their ``Bayesian'' content sharpened
just as in the quantum case) 
by saying that for every choice of an 
Alice and Bob measurement, the joint distribution it determines 
conditionalizes in the usual Bayesian (classical probabilistic)
manner, according to (\ref{cbayes}).  And, just as in the quantum
case this held for joint states of distinct systems, here it 
holds whenever the coupling between two systems does not
permit signalling.  The relation between the simultaneous
``Bayesian'' nature of the conditioning (on a result of
Alice) of the probabilities of
{\em all} of  Bob's measurements, and no-signalling in the Bob-to-Alice
direction, is essentially just a restatement
of Theorem 3.4.  It dramatizes the ``least-disturbance'' nature
of the nearest-to-Bayesian updating rule: other rules correspond
a ``more serious'' or ``more physical'' disturbance, in the sense
that the disturbance itself could be used to carry information (signal).  

The close connection of conditioning of that occurs in the FNS tensor
product, and Bayesian updating, was also noted by Randall and Foulis,
who give an ``Operational Bayes Theorem:'' 

\begin{theorem}(\cite{Foulis-Randall80b}, Theorem 2.6)
Let $\omega$ be a state on $(XY, \fa\fb)$, $a \in X$, $b \in Y$, $F \in \fb$.  Then
\beq
\omega_a(b) = \omega_B(b) \omega_b(a)/(\sum_{c \in F} \omega_B(a)\omega_c(a))\;.
\eeq
\end{theorem}

\textitem{Influence-Free States Linearized} The function $y \mapsto \omega(Ey)$
is independent of $E \in {\frak A}$ if and only if, for every fixed $y
\in Y$, the map $\omega_{y} : x \mapsto \omega(xy)$ is a
(non-normalized, but positive) {\em weight} on $(X,{\frak A})$.  If it
is also the case that $x \mapsto \omega(xF)$ is independent of $F \in
{\frak B}$ -- that is, if $\omega$ is influence-free -- then the map $y
\mapsto \omega_{y}$ can be interpreted as a vector-valued weight on
$(Y,{\frak B})$ with values in the space $V(X,{\frak A})$ of ``signed
weights" (i.e., linear combinations of states) on $(X,{\frak A})$.
         
If $V(X,{\frak A})$ and $V(Y,{\frak B})$ are finite-dimensional, we
can blithely dualize the foregoing picture: every influence-free weight on
$(X,{\frak A}) \times (Y,{\frak B})$ determines, and is determined by,
a positive linear operator $\widehat{\omega} : V^{\ast}(Y,{\frak B})
\rightarrow V(X,{\frak A})$ with the property that $\omega(1) \in
\Omega(X,{\frak A})$. More generally, given an (unnormalized) positive
influence-free weight $\omega$, we obtain a positive map $\widehat{\omega}
: V^{\ast}(Y,{\frak B}) \rightarrow V(X,{\frak A})$, and any such
map $\phi$, conversely, determines a influence-free weight $\omega$ via
$\phi(f_{x}) = \omega(x)$ for all $x \in X$. Thus, we have

\begin{theorem} The map $\omega \mapsto \widehat{\omega}$ is an 
affine isomorphism between the cone of positive influence-free weights on ${\frak A}  \times {\frak B}$ and the cone of positive linear maps from 
$V^{\ast}(Y,{\frak B})$ to $V(X,{\frak A})$.
\end{theorem}

\begin{example} Specializing to the (finite-dimensional) Hilbert space 
setting, this gives us, for every influence-free state on ${\frak F}(\H) \times {\frak F}(\K)$, a 
positive linear map (a superoperator) 
\[\widehat{\omega} : {\cal L}_{sa}(\H) 
\rightarrow {\cal L}_{sa}(\K)\] 
on the space of bounded self-adjoint 
operators on $\H$, satisfying $\tr(\phi(\1)) = 1$.  This extends, (via the cartesian 
decomposition) to a positive linear map on the full space ${\cal L}(\H)$. 
Conversely, any positive linear map $\phi : {\cal L}(\H) \rightarrow 
{\cal L}(\H)$ determines a state $\omega$ on ${\frak F}(\H) \times {\frak 
F}(\K)$ via \[\omega(xy) := \tr[\phi(\proj{x}) \proj{y})] \equiv
\dmelement{y}{\phi(\proj{x})}\;.\]
where $\proj{x}$ is the orthogonal projection 
operator determined by $x \in \H$.   

Thus, the set of influence-free states on ${\frak F}(\H) \times {\frak F}(\K)$ is 
affinely isomorphic to the space of positive linear maps on ${\cal 
L}(\K)$.\footnote{All of this pretty much goes through even in the 
infinite-dimensional setting, as long as ${\frak A}$ and ${\frak B}$ are 
locally countable. \cite{Wilce92}.}\\
\end{example} 

\section{Operator Representations of Influence-Free States}

We now specialize to the case in which $\dim(\H) = \dim(\K)$. For
simplicity, we assume that $\H = \K$. In this setting, one can
represent influence-free states on ${\frak F}(\H) \times {\frak
F}(\K)$ by operators on $\H \otimes \H$, using the following useful result:

\begin{proposition}(Folklore) \label{prop: folklorico}
For any linear map $\phi : {\cal L}(\H) \rightarrow {\cal L}(\H)$,
there exists a unique operator $W = W_{\phi}$ on $\H \otimes \H$ such
that, for all $x,y,u,v \in \H$,
\begin{boxx} \dmelement{y}{\phi(\proj{x})} \ 
= \ \bra{y}\bra{x} W \ket{x}\ket{y}.\end{boxx} Conversely, every
operator $W$ on $\H \otimes \H$ arises in this way from a unique
linear map $\phi: {\cal L}(\H) \rightarrow {\cal L}(\H)$.
\footnote{In a suitable basis, the matrix for $W_{\phi}$ is just the
so-called Choi matrix for $\phi$. This is discussed below.}
\end{proposition} 

{\bf Proof:} For any linear operator on ${\cal L}(\H)$, the quantity 
$\bra{y}\phi(\proj{x}) \ket{y}$ is bi-quadratic in $x$ and $y$. 
Polarizing twice, see that $\phi$ is uniquely determined by the form 
\[(x,u,y,v) \mapsto \bra{v} \phi(\outerp{x}{u})\ket{y}.\]
Note that this is linear in $x$ and $y$, conjugate-linear in $u$ and $v$. 
Accordingly, there is a unique {\em sesquilinear} form
$\Phi$ on $\H \otimes \H$ satisfying
\[ \Phi(\ket{x}\ket{y}, \ket{u}\ket{v}) 
\ := \  \bra{v} \phi(\outerp{x}{u})\ket{y}.\]
By the Riesz representation theorem (cf. \cite{Kadison-Ringrose}, Theorem 2.3.1;
cf. also Theorem 2.4.1 for a result close to Proposition \ref{prop: folklorico}), 
there is a 
unique operator $W$ such that $\Phi( \tau_1, \tau_2) = 
\matelement{\tau_2}{W}{\tau_1}$ for all tensors $\tau_1, \tau_2 \in \H \otimes \H$. Setting 
$\tau_1 = \ket{x}\ket{y}$ and $\tau_2 = \ket{u}\ket{v}$ gives the result. \QED 

This immediately yields the following ``unentangled Gleason theorem":
             
\begin{corollary} (\cite{Klay-Randall-Foulis}; see also 
\cite{Fuchs2001a, Fuchs2002a, Wallach}): Let $\H$ be a finite-dimensional complex Hilbert space. For every influence-free state $\omega$ on 
${\frak F}(\H) \times {\frak F}(\H)$, there exists a self-adjoint
operator $W$ on $\H$ such that $\omega(xy) \ = \bra{y}\bra{x} W
\ket{x}\ket{y} $ for all unit vectors $x, y \in \H$.
\end{corollary}
    
Evidently, the operator $W$ must be {\em positive on pure tensors} 
(POPT), in that $\bra{y}\bra{x} W \ket{x}\ket{y} \geq 0$ for 
all $x, y \in \H$. However, $W$ need not be positive:
 
\begin{example} Let $S(\ket{x}\ket{y}) = \ket{x}\ket{y}$, i.e., $S$ is the 
unitary ``swap" operator. Then $S$ is POPT, since 
$\bra{y}\bra{x} S \ket{x}\ket{y} = 
\bra{x}\bra{y}, \ket{x}\ket{y} \rangle = \inner{x}{y}\inner{y}{x} 
= |\inner{x}{y}|^2$.
But $S$ is certainly not positive. Indeed, if $\tau = \ket{x}\ket{y} - 
\ket{y}\ket{x}$, then $S\tau = -\tau$, whence $\dmelement{\tau}{W}= -
 \|\tau\|^2$.  
\end{example}
            
The question now arises: When is the POPT operator $W_\phi$ arising
from a positive linear map $\phi : {\cal L}(\H) \rightarrow {\cal
L}(\H)$ in fact {\em positive} on $\H \otimes \H$?

Recall that a linear map $\phi : {\cal L}(\H) \rightarrow {\cal
L}(\H)$ is {\em completely positive} (CP) iff the map $\phi
\otimes \Id : {\cal L}(\H \otimes \K) \rightarrow {\cal L}(\H \otimes
\K)$ remains positive for all Hilbert spaces $\K$. It is a standard
result (due independently to Hellwig and Kraus (\cite{Hellwig69a},
\cite{Hellwig70a}; \cite{ Kraus}) and to Choi
\cite{Choi}) that $\phi$ is CP if and only if it can be expressed in the form
\[\phi(X) = \sum_{i} A_{i} X {A}^\dagger_{i}\] for operators $A_{i}$ on
$\H$. Such a decomposition of $\phi$ is called a {\em Hellwig-Kraus
representation}, or HK representation for short.  (Many authors in 
quantum information theory, probably through familiarity with \cite{Kraus}
rather than \cite{Hellwig69a, Hellwig70a}, just call it a Kraus decomposition.)

\begin{theorem}{(Choi, Hellwig and Kraus)} \label{CHK theorem}
Let $W = W_{\phi}$ be the operator associated 
with the linear map $\phi : {\cal L}(\H) \rightarrow {\cal L}(\H)$
as in Proposition 4.1. Then $W$ is positive iff $\phi$ is completely
positive.
\end{theorem}


\textitem{Matrix Representations} Suppose now that $E = \{e_{i}\}$ is
an ordered orthonormal basis for $\H$. We can then represent a map
$\phi : {\cal L}(\H) \rightarrow {\cal L}(\H)$ by an operator-valued
matrix $\Phi_{i,j} := \phi(\outerp{e_i}{e_j})$. If we represent each
of the entries $\Phi_{i,j}$ by an $n$-by-$n$ matrix relative to the
same basis $E$, we find that the $(i,j)-(k,l)$-th entry is
$\Phi_{i,j,k,l} = \bra{e_l}\phi(\outerp{e_i}{e_k}) \ket{e_j}$. This is
often called the {\em Choi matrix} for $\phi$. We'll write $\Ch(\phi)$
for this matrix (taking the basis $\{e_i\}$ as understood).  For the
$i,j$ block of it, we'll write ${\rm Ch}\phi)^{ij}$; \\

\textitem{Representation Using a Maximally Entangled Pure State} Again
let $E$ be an orthonormal basis for $\H$. The product basis
$\{\ket{a}\ket{b}| a, b \in E\}$ yields an operator basis for ${\cal
L}(\H \otimes \H)$ consisting of the operators
\[\ket{a}\ket{b} \bra{c}\bra{d} \ = \ \outerp{a}{d} \otimes \outerp{b}{c},\]
$a,b,c$ and $d$ running over $E$. By expanding the operator 
$W_\phi$ defined in Proposition \ref{prop: folklorico} in this basis, one can 
show that 
\begin{boxx} W_\phi \ = \ (\Id \otimes \phi)(T) \end{boxx}
where $T$ is the (unnormalized) 
pure maximally entangled state given by \[T \ = \ \sum_{a,c \in E} \outerp{a}{c} \otimes \outerp{a}{c}.\] \\

\section{Decomposable States} 

The structure of the full set of positive maps between ${\cal L}(\H)$ and 
${\cal L}(\K)$ is very complicated, even in low dimensions.  A set of 
maps larger than the set of CP maps, but still tractable, is that of
{\em decomposable} maps. If $\phi$ is a CP map, then the map 
\[\tp{\phi}: X \mapsto \phi(X^{t})\]
obtained by composing $\phi$ with the a transposition map is said to be 
{\em co-completely positive} (co-CP). A map of the form $\phi = \psi + 
\eta$ where $\psi$ is CP and $\eta$ is co-CP is said to be {\em 
decomposable}. The set of CP maps, the set of co-CP maps, and the set 
of decomposable maps are all convex cones in the space of linear 
operators on ${\cal L}(\H)$, with the cone of decomposable maps being the 
convex span of the CP and co-CP cones. 

We wish to understand the extremal structure of the cone of decomposable 
maps. By an {\em extremal point} of a cone, we mean a point generating an 
extreme ray. Evidently, any extremal point of the convex span of two cones 
must be extremal in one of the two cones to begin with; in general, however, 
some extremal points of the original cones will be ``swallowed up" in the 
passage to  the convex span. Our aim here is to prove that this {\em doesn't} 
happen here: i.e., extreme CP maps and extreme co-CP maps all remain extreme 
in the larger cone of decomposable maps.      

{\bf Notation:} Given an operator $A$ on $\H$, we'll write $\phi_{A}$ for the CP map 
\[\phi_{A} : X \mapsto AXA^\dagger.\]
Evidently, any extreme CP map is of this form.   Notice that $\phi_{A}\phi_{B} = \phi_{AB}$. This gives us the trivial but useful 
        
\begin{lemma} Let $\psi : {\cal L}(\H) \rightarrow {\cal L}(\H)$ be CP. 
Then for any operator $A$ on $\H$, $\phi_{A} \circ \psi$ and $\psi \circ 
\phi_{A}$ are likewise CP maps. 
\end{lemma}

\textitem{Transpositions} 
Let $\M_{n}$ denote the $\ast$-algebra of $n
\times n$ complex matrices. Any orthonormal basis $E = \{\ket{e_i}_i\}$
for $\H$ induces an isomorphism ${\cal L}(\H) \rightarrow \M_{n}$
given by $X \mapsto [X]^{E}$, where $[X]^{E}_{ij} =
\matelement{e_i}{X}{e_j}$.  Accordingly, the map $M \mapsto M^{t}$
pulls back to an anti-linear map $J : {\cal L}(\H) \rightarrow
{\cal L}(\H)$ given by $[J(X)] = [X]^{t}$.  Let $\sigma$ be the
anti-linear operator on ${\cal L}(\H)$ corresponding to transposition
with respect to another orthonormal basis $F = \{\ket{f_j}\}$ for
$\H$. Let $U$ be the unitary map sending $\ket{e_i}$ to $\ket{f_j}$, so
that $[X]^{F} = \bra{{e_j}}U^\dagger XU\ket{e_i}=
[U^{\dagger} X U]^{E}$. Thus,
\[\sigma(X) = J(UXU^{\dagger}) = J(\phi_{U}(X)).\]

\begin{lemma} \label{lemma: composition with CP}
Let $\psi$ be a linear map ${\cal L}(\H) \rightarrow 
{\cal L}(\H)$. If  $\psi \circ J$ is CP,  then so is $\psi \circ \sigma$.
\end{lemma}

{\bf Proof:} By the foregoing discussion, $\psi \circ \sigma = \psi \circ J 
\circ \phi_{U}$.  By assumption, $\psi \circ J$ is CP. By Lemma A, so is 
$(\psi \circ J) \circ \phi_{U}$. $\Box$ 

It follows that $\psi$ is co-CP iff $\psi \circ J$ is CP for {\em any} 
transposition $J$. As in the introduction, we'll use the notation $X^{t}$ 
for the operator $J(X)$, where $J$ is some fixed but unspecified 
transposition; we'll also write $\tp{\phi}$ for the map $\phi \circ J : 
X \mapsto \phi(X^{t})$.  
                             
\begin{lemma} \label{lemma: composition with co-CP}
Let $\psi$ be co-CP. Then for any operator $A$ on $\H$, 
$\phi_{A}\circ \psi$ is again co-CP.
\end{lemma}

{\bf Proof:} By assumption, $\psi^{t} = \psi \circ J$ is CP. Now, $(\phi_{A} 
\circ \psi) \circ J = \phi_{A} \circ (\psi \circ J)$. By Lemma 5.1, this 
is also CP. Hence, $\phi_{A} \circ \psi$ is co-CP. $\Box$ 
 
The following lemma gathers some elementary but helpful facts about Choi 
matrices. As above, $\{\ket{e_i}\}$ is a fixed orthonormal basis for $\H$ and 
$\tp{\phi} = \phi \circ J$ where $J$ is transposition with respect to 
this basis. 

\begin{lemma} \label{lemma: stuff about Choi of extremal CP and co-CP}
Let $A, B \in {\cal L}(\H)$, with $A\ket{e_i} = \ket{a_i}$, and let 
$\phi : {\cal L}(\H) \rightarrow {\cal L}(\H)$ be any linear map. Then 
\begin{mlist}
\item[(a)] $\Ch(\phi_{A} + \phi_{B}) = \Ch(\phi_A) + \Ch(\phi_B)$; 
\item[(b)] $\Ch(\phi_{A})^{ij} = \outerp{a_i}{a_j}$. 
\item[(c)] $\Ch(\tp{\phi}_{A}) = \outerp{a_j}{a_i}$.
\item[(d)] $\Ch(\tp{\phi}) = \Ch(\phi)^{\idop \otimes t}$.
\end{mlist}
\end{lemma}

Here $X^{\idop \otimes t}$ is the {\em partial transpose} of $X$, i.e.
the bipartite state $X$ subjected to the extension, by the identity on map
the one factor, of the transpose map acting on the other factor.

\begin{lemma} \label{lemma: about block-diagonal Choi matrices}
Suppose that $\Ch(\phi)$ is block-diagonal. Then $\phi$ 
is CP iff $\phi$ is co-CP. 
\end{lemma}

{\bf Proof:}  The assumption is that $\phi(\outerp{e_i}{e_j}) = 0$ for $i \not = j$. 
It follows that $\tp{\phi}(\outerp{e_i}{e_j}) = 
\phi(\outerp{e_j}{e_i}) = 
\phi(\outerp{e_i}{e_j})$ for all $i, j$, i.e, $\Ch(\tp{\phi}) = \Ch(\phi)$. Now 
invoke Theorem \ref{CHK theorem}. $\QED$

We now prove the advertised result, which we restate for convenience: 

\begin{theorem} \label{theorem: extremal decomposable}
Let $\phi$ be an extremal map in the cone of CP maps, or 
an extremal map in the cone of co-CP maps. Then 
$\phi$ is extremal in the cone of decomposable maps. 
\end{theorem}

Equivalently, the extremal quantum states and the extremal PPT states 
(states corresponding to Hermitian operators with positive partial transpose)
remain extremal in the cone of decomposable states.

Note that the map $\phi \mapsto \tp{\phi}$ is an affine isomorphism
from the cone of CP maps to that of co-CP maps.  Thus, without loss of
generality we can focus on an extreme CP map \[\phi : X \mapsto
AXA^{\dagger}\] and ask whether this can be expressed nontrivially as
a sum of CP and co-CP maps. We therefore suppose in what follows that
$\phi \ = \ \psi + \eta$ where $\psi$ and $\eta$ are respectively
non-zero completely and co-completely positive maps with Hellwig-Kraus
representations \[\psi = \sum_k \phi_{B_k} \ \ \text{and} \ \ \eta =
\sum_l \tp{\phi}_{C_l}.\] Our aim is to prove that $\psi$ and $\eta$
are in fact multiples of $\phi$.  The strategy will be to show that
$\eta$ lies in the separable cone, i.e., is both CP and co-CP: the
extremality of $\phi$ in the CP cone then yields the desired
result. The key observation (along with Lemma 
\ref{lemma: about block-diagonal Choi matrices}) is the following:

\begin{lemma}
\label{big lemma}
Let $\phi, \psi, \eta$ be as above.  That is,
$\phi = \psi + \eta$ with $\phi = \phi_A$, $\psi = \sum_k \phi_{B_k},
\eta = \sum_l \phi_{C_l}^t.$ so that $\phi$ is CP
and $\eta$ co-CP. Suppose 
$\{\ket{e_i}\}$ is an orthonormal 
basis for $\H$, and let $\ket{a_i} = A\ket{e_i}$.  The 
Choi matrix $\Ch(\eta)$ for $\eta$ relative to the basis $\{\ket{e_i}\}_i$
has the following 
form: For all $i,j$, 
\begin{mlist}
\item[(a)] if $a_i$ and $a_j$ are linearly dependent, then 
$\Ch(\eta)^{ij} = z_{ij} \outerp{a_i}{ a_i}$ for some complex coefficient 
$z_{ij}$; 
\item[(b)] if 
$a_i$ and $a_j$ are linearly independent, then $\Ch(\eta)^{ij} = 0$. 
\end{mlist}
\end{lemma}

{\bf Proof:} (a) Let $\ket{b_{ki}} := B_{k}\ket{e_i}$, and 
$\ket{c_{li}} = C_{l}\ket{\ket{e_i}}$. Since $\phi 
= \psi + \eta$, we have $\Ch(\phi) = \Ch(\psi) + \Ch(\eta)$. Invoking Lemma 
\ref{lemma: stuff about Choi of extremal CP and co-CP}, 
we obtain \[\outerp{a_i}{a_j} \ = \ \sum_k \outerp{b_{ki}}{b_{kj}} + \sum_l 
\outerp{c_{lj}}{c_{li}}\] for all $i,j$. With $i = j$, we have 
\[\outerp{a_i}{a_i} 
\ =  \ \sum_{k} \outerp{b_{ki}}{b_{ki}} + \sum_{l} \outerp{c_{li}}{c_{li}}.\] 
Now, $\outerp{a_i}{a_i}$, $\outerp{b_{ki}}{b_{ki}}$ 
and $\outerp{c_{li}}{c_{li}}$ 
are non-negative multiples of one-dimensional orthogonal projections; these 
are extremal in the full cone of positive operators, so we must have 
$\ket{b_{ki}} 
= b_{ki} \ket{a_i}$ and $\ket{c_{li}} = c_{li}\ket{a_i}$ 
for all $k, l$, where $b_{ki}$ and 
$c_{li}$ are complex scalars. In particular, then, \[\Ch(\psi)^{ij} = r_{ij} 
\outerp{a_i}{a_j} \ \text{and} \ \Ch(\eta)^{ij} = s_{ij} \outerp{a_j}{a_i}\] where \[r_{ij}  = \left ( \sum_{k} b_{ki}\overline{b}_{kj} \right ) 
\ \text{and} \ s_{ij} = \left (\sum_{l} c_{lj}\overline{c}_{li} \right ).\] 
Suppose now that $a_{i}$ and $a_{j}$ are linearly dependent, i.e, $a_{j} = k_{ji} a_{i}$ for some non-zero $k_{ji}$. Then $\Ch(\eta)^{ij} = z_{ij} 
\proj{a_i}$ where $z_{ij} = s_{ij} k_{ji}$. This yields part (a). If 
$a_{i}$ and $a_{j}$ are independent, then so are $\outerp{a_i}{a_j}$ and 
$\outerp{a_j}{a_i}$. From the fact that \[\outerp{a_i}{a_j} = r_{ij} 
\outerp{a_i}{a_j} + s_{ij} \outerp{a_j}{a_i}.\] we see that  $s_{ij} = 0$, i.e., $\Ch(\eta)^{ij} = 0$. $\Box$ 

{\bf Proof of Theorem~\ref{theorem: extremal decomposable}:} 
We first consider the case in which $A$ is self-adjoint.  
Choose the orthonormal basis $\{\ket{e_i}\}$ of Lemma
\ref{big lemma} so as to 
diagonalize $A$. In this case, $\ket{a_i}$ and $\ket{a_j}$ 
are linearly dependent only 
if one of them is zero, or if $i = j$. By Lemma \ref{big lemma}, 
the only cases in which 
$\Ch(\eta)^{ij} \not = 0$ are those in which $i = j$, i.e., $\Ch(\eta)$ is 
itself block-diagonal. Thus, by Lemma 
\ref{lemma: about block-diagonal Choi matrices}, $\eta$ is separable.  

Suppose now that $A$ is arbitrary. By the polar decomposition theorem
(c.f. e.g. \cite{Horn-Johnson}, Theorem 7.3.2),
there exists a partial isometry $W$ such that $|A| = WA$ and
$W^{\dagger}|A| = A$.  ($|A|$ is defined as $\sqrt{A^\dagger A}$, while
a {\em partial isometry} is defined as an operator $W$ such that 
$WW^\dagger$ (and hence also $W^\dagger W$) is a projection.)
Thus, \[\phi_{|A|} = \phi_{W}\circ \phi_{A} =
\phi_{W}\circ \psi + \phi_{W}\circ \eta.\] By Lemmas \ref{lemma:
composition with CP} and \ref{lemma: composition with co-CP},
$\phi_{W}\circ \psi$ and $\phi_{W} \circ \eta$ are respectively CP and
co-CP. Hence, $|A|$ satisfies the same hypotheses as $A$.  In
particular, we can invoke the preceding argument to conclude that
$\phi_{|A|}$ is separable.  But then we can apply Lemmas \ref{lemma:
composition with CP} and \ref{lemma: composition with co-CP} again to
conclude that $\phi_{W^{\dagger}}\circ \phi_{|A|} = \phi_{A}$ is also
separable.  $\Box$

{\em Remark:} It is a classical result of Choi \cite{Choi} that if
$\dim(\H) = 2$, all positive linear maps ${\cal L}(\H) \rightarrow
{\cal L}(\H)$ are decomposable.\\
 

\section{Teleportation and POPT States:  Difficulties with 
combining} \\
{\bf \underline{free-no-signalling and quantum composition of systems}}
 
Here we consider what happens if we view the Foulis-Randall
FNS
product as the way to couple quantum systems distant from each other, but 
attempt to couple the different ``local'' systems with the ordinary
quantum-mechanical product (reasoning that locally, measurements
with ``entangled outcomes'' are possible).   We may view the situation 
as similar to that in many quantum protocols illustrating ``nonlocal'' effects
associated with entangled states, or distilling entanglement, etc..:  \
two agents distant from each other, ``Alice'' and ``Bob,''
are viewed as each having a number of systems, say $A_1... A_n$, $B_1....B_n$,
under their control.  We show in this section that attempts to mix these 
two types of coupling lead to pathologies.  In particular, the following
is usually
considered a {\em desideratum} for a notion of compound system $AB$ 
composed of two subsystems $A$ and $B$.

\textitem{Desideratum}  For every pair of states $\omega$ on $A$,
$\lambda$ on $B$,
there exists a ``product'' $(\omega \lambda)$
of these two states on the compound system,
such that in that state, the probability of the pair of outcomes $xy$
is the product of its probabilities under the pair of subsystem states:
$(\omega \lambda)(xy) = \omega(x) \lambda(y)$.  

We consider $A_1$ and $A_2$ to be coupled quantum-mechanically, 
and similarly with $B_1$ and $B_2$.  We then couple the quantum
mechanical test-space $A_1 \otimes A_2$
with $B_1 \otimes B_2$, via the FNS tensor product, obtaining
a test space we will call $AB$.  We also consider
the  test-spaces, and states, obtained by coupling $A_1$ with $B_1$,
and $A_2$ with $B_2$, via the FNS tensor product (call the resulting
test spaces
$1$ and $2$).  We ask if we 
can view the test space $AB$ as any reasonable kind of product of 
$1$ and $2$.  

\begin{proposition}\label{ivory}
There exist pairs of states on 
$1$ and $2$ whose product is not a state of $AB$.  
\end{proposition}

That is, the two types of coupling, ``local quantum mechanical'' and
``nonlocal free-no-signalling (FNS)'', cannot be combined in a manner
consistent with Desideratum 1.  The intuitive argument is simple: if
Alice and Bob share both a Bell state and a POPT state that is
nonpositive, Alice may use the Bell state and a local measurement in
an entangled basis of Bell states to teleport her part of the POPT
state to Bob; since local coupling is quantum-mechanical, Bob has
available measurements with ``entangled'' outcomes, which is
inconsistent with his possessing both parts of a POPT state.  We now
formalize this argument, and explain why it proves Proposition
\ref{ivory}.

The setting here is similar to that of ordinary quantum teleportation
protocols, except that we suppose that {\em both} Bob's and Alice's
systems are bipartite. Thus, we have $\H_{\rm Alice} \equiv
\H_A := \H_{A_1} \otimes
\H_{A_2} \ \ \text{and} \ \ \H_{\rm Bob} \equiv H_B 
:= \H_{B_2} \otimes \H_{B_1}$, We assume
here that all four Hilbert spaces $\H_{A_i},\H_{B_i}$ are copies of a
common finite-dimensional Hilbert space $\H$ of (finite) dimension
$n$.  ``Copies of a common space'' rather than just ``isomorphic Hilbert
spaces'' implies we have selected a commuting set of isomorphisms between
them;  equivalently, we have selected and identified a ``standard'' orthonormal
basis in each.  This means that it makes sense to speak of a given operator
$O$, acting on different systems.  (The alert reader may wonder why we defined $\H_B$ as 
$\H_{B_2} \otimes \H_{B_1}$, rather than with the more natural 
ordering $\H_{B_1} \otimes \H_{B_2}$ used in the informal discussion
above.\footnote{Other readers are urged to treat themselves to a double 
cappucino--or a mat\'e--before continuing;  a chai latt{\'e} might do,
if you insist, though we cannot be held responsible for the consequences
of routinely indulging in such beverages, which we have heard may end 
in walking up and down Telegraph Avenue clad in nothing but a poncho, beads,
and patchouli oil.}  
The reason is that if we had
used the other ordering, we would have had to replace the second
occurence of the state $W$ in the first line of Theorem \ref{bleah}
with $\tilde{W}$, defined as ${\rm Swap} ~W~ {\rm Swap}^\dagger$, where
Swap is the unitary operator that swaps the states of systems $B_1$ 
and $B_2$.  In other words, the standard teleportation protocol where
Alice measures in a Bell basis and tells Bob her result, 
and the ``source'' system $\H_{A_1}$ is entangled with $\H_{B_1}$ so
that $\H_{A_1} \otimes \H_{B_1}$ is in state $W$, ends up, in the
case where the Bell measurement gives the standard maximally entangled
state $T$ (defined below), ``pivoting'' $\H_{A_1}$'s entanglement into $\H_{B_2}$, where
we can think of it as ``pivoting'' around $\H_{B_1}$, whose role (and
state) stays constant.)

The canonical isomorphisms between Hilbert spaces given above also
imply that it makes sense to speak of a given operator defined on the
(ordered) tensor product of two systems, as acting on some other
(ordered) tensor product.  The convention we will use here is that
(given commuting canonical isomorphisms between Hilbert spaces
$\H,\J,\K,\L$) if some operator $W$ is specified as an operator on $\H
\otimes \J$, then $W$ on a different tensor product $\K \otimes \L$ is
determined by the requirement that its matrix elements in the tensor
product basis $\ket{i}^{\K}\ket{j}^{\L}$ for $\K \otimes \L$ are the
same as its elements in the tensor product basis
$\ket{i}^{\H}\ket{j}^{\J}$ for $\H \otimes \J$.  Here $\ket{i}^{\X}$
is the standard basis vector $\ket{i}$ for system $\X$.  Note that the order
of the tensor product matters: $W$ acting on $\K \otimes \L$ is the
same as ${\rm Swap}~W~{\rm Swap}^\dagger$ acting on $\L \otimes \K$.
We use several notations to help specify which of a set of
canonically isomorphic systems an 
operator acts on, or a basis element belongs to: one is to use a
superscript specifying the system, as we did above with basis vectors;
another, for tensor products of operators, is to put a subscript on the
tensor product sign used between the operators, related to subscripts
used to identify Hilbert spaces:  Thus, for example,
$X \otimes_{12} Y$ would be interpreted as acting on $\H_1 \otimes \H_2$ (and
similarly $X \otimes_{AB} Y$ acts on $\H_A \otimes \H_B$.    

As usual, we think of $\H_{A}$ and $\H_{B}$ as representing
 spatially separated, ``local" subsystems of the total system
 $\H_{Total} := \H_{A} \otimes \H_{B}$. There are also two {\em
 non-local} subsystems of interest to us, namely, $\H_1 := \H_{A_1}
 \otimes \H_{B_1}$ and $\H_2 := \H_{A_2} \otimes \H_{B_2}$.  We assume
 that $H_1$ is in a state represented by an operator $W$ and that
 $\H_2$  is in a maximally entangled pure state represented
 (up to normalization) by the rank-one projection 
\[T \ = \
 \frac{1}{n} \sum_{e,f \in E} \ket{e}\ket{e}\bra{f}\bra{f} =
\frac{1}{n} 
\sum_{e,f \in E} \outerp{e}{f} \otimes_{A_2,B_2} \outerp{e}{f},\] where $E$ is some
 fixed (but arbitrary) orthonormal basis for $\H_{A_2} \simeq \H_{B_2}$.  
(Note that it was not strictly necessary to use the subscript $A_1A_2$ on the tensor
product sign on the right, since we identified the overall system as $\H_2$ and the ordering
of $A_1, A_2$ was specified in the definition of $\H_2$.  We will sometimes, but not
always, leave 
subscripts off of tensor product signs when the system is otherwise identified.
Associativity of the ordered tensor product will also be used without further ado.)
The
 state of the total system is thus represented by $W \otimes_{12} T$.
In the protocol, Alice makes a measurement on $\H_A$,
having $T$ (i.e. $T^A$) as a possible outcome.  If we calculate
the state that the projection postulate would describe for the system,
conditional on that measurement outcome, we obtain, 
up to normalization by $\alpha :=
 \tr((T \otimes_{AB} \1)(W \otimes_{12} T))$, \[(T \otimes_{AB} \1)(W \otimes_{12}
 T)(T \otimes_{AB} \1).\] Similarly, if Bob obtains $T$ as the outcome of
 a measurement on his system, then the final total state will be, up
 to normalization,
\[(\1 \otimes_{AB} T)(W \otimes_{12} T)(\1 \otimes_{AB} T).\]
 
\begin{theorem} \label{bleah}
Let $T$ and $U$ be as described above. Then for 
any operator $W$ on $\H_1$,  
\begin{boxx}
\begin{array}{c}
(T \otimes_{AB} \1)(W \otimes_{12} T)(T \otimes_{AB} \1) = \alpha T \otimes_{AB} W.\\
\text{and}\\
(\1 \otimes_{AB} T)(W \otimes_{12} T)(\1 \otimes_{AB} T) = \alpha W \otimes_{AB} T.\end{array}
\end{boxx}
\end{theorem}

{\em Remark:} If in the preceding formula we replace $\H_{B_1}$ by the
one-dimensional Hilbert space ${\Bbb C}$, so that $\H_{B} = \H_{B_2}$,
then we recover the usual teleportation scheme.  The present scheme makes
explicit the fact (which has been observed and exploited many times before) 
that the standard teleportation protocol serves to 
teleport not only a state, but its entanglement with a
system not otherwise involved in the protocol.

It will be convenient to work with the ``un-normalized state"
\[Q = n T = \sum_{e,f \in E} \outerp{e}{f} \otimes \outerp{e}{f},
\] instead of $T$. We'll normalize
later. We'll make use of the following

\begin{lemma} Let $E$ be an orthonormal basis for $\H = \H_{A_2}
= \H_{B_2}$, as above. For any operator of the form $A = \outerp{x}{y}
\otimes \outerp{u}{v}$, with $x,y,u,v \in E$, \[ QAQ = \left \{
\begin{array}{ll} Q & \text{if} \ \ x = u \ \text{and} \ y = v;\\ 0 &
\text{otherwise} \end{array} \right . \]
\end{lemma}

{\bf Proof:} Direct computation yields
\begin{eqnarray*}
Q A Q & = & \sum_{e,f,e'f'} (\outerp{e}{f} \otimes \outerp{e}{f})
(\outerp{x}{y} \otimes \outerp{u}{v}) 
(\outerp{e'}{f'} \otimes \outerp{e'}{f'})\\
& = & \sum_{e,f,e',f'} (\outerp{e}{f})(\outerp{x}{y})(\outerp{e'}{f'}) 
\otimes (\outerp{e}{f})(\outerp{u}{v})(\outerp{e'}{f'})\\
 & = & \sum_{e,f,e',f'} \inner{f}{x}\inner{y}{e'}
\inner{f}{u}\inner{v}{e'}
\outerp{e}{f'} \otimes \outerp{e}{f'} 
\end{eqnarray*}
The inner products here are zero except where $x = f = u$ and $y = e'
= v$. In other words, for the result to be non-zero, the input vector
$A$ must have the form $A = \outerp{x}{y} \otimes \outerp{x}{y}$,
and in this case we do in fact get $QAQ = Q$. $\Box$ \\

{\bf Proof of Theorem~\ref{bleah}:} The product basis 
$E \otimes E = \{\ket{a}\ket{b} | a, b \in E\}$ gives us an 
operator basis for ${\cal L}(\H_1)$, namely, 
\[\{ \ket{a}\ket{b} \bra{c}\bra{c} = \outerp{a}{d} \otimes \outerp{b}{c} 
: a,b,c,d \in  E\}.\] 
 Expanding $W$ in this basis, we have 
 \[W \ = \ \sum_{a,b,c,d} W_{a,b,c,d} \outerp{a}{d} \otimes \outerp{b}{c}\]
Thus,
\[(W \otimes_{12} Q) \ = \ \
\sum_{a,b,c,d,e,f} W_{a,b,c,d} \outerp{a}{d} 
\otimes \outerp{e}{f} \otimes \outerp{e}{f} \otimes \outerp{b}{c}.\]
Applying $Q \otimes_{AB} \1$ to both sides, we have 
\[(Q \otimes_{AB} \1) U (W \otimes_{12} Q) U (Q \otimes_{AB} \1) =  \sum_{a,b,c,d,e,f} W_{a,b,c,d} 
Q (\outerp{a}{d} \otimes \outerp{e}{f}) Q \otimes_{AB} (\outerp{e}{f} \otimes \outerp{b}{c}).\]
According to our Lemma, we obtain non-zero terms only where
$\outerp{a}{d} = \outerp{e}{f}$, i.e., where $e = a$ and $f = d$, and
in this case we have $Q(\outerp{a}{d} \otimes \outerp{a}{d}) Q =
Q$. Thus, we end up with
\[\sum_{a,b,c,d} Q \otimes_{AB} (W_{a,b,c,d} \outerp{a}{d}
\otimes \outerp{b}{c}) = Q \otimes_{AB} W.\]
Substituting $n T$ for $Q$ throughout yields the fist line in the
desired result (boxed equations in Theorem \ref{bleah};  
the proof of the second line is entirely analogous.
$\Box$
 
{\bf Remark:} In this version of teleportation, Bob has some access to
the state to be teleported, through $W$'s marginal on $\H_{B_1}$. Note
that this remains unchanged after teleportation.  (Indeed, whether we
place $W$ on $\H_1 \equiv \H_{A_1} \otimes \H_{B_1}$ 
or on $\H_{A_1} \otimes \H_{B_2}$, teleportation always
``pivots" $W$ about whichever component of Bob's system partakes of
$W$. In particular, if we start with $W$ on $\H_{A_1}\otimes \H_{B_2}$, 
(and switch $T$ compatibly as well) 
we end up, not with $W$ on $\H_{B_2} \otimes \H_{B_1}$, 
but with $W$ on $\H_{B_1} \otimes \H_{B_2}$. )

We have only shown what happens conditional on 
a particular outcome of Alice's measurement, the standard entangled
state $T$.  It is straightforward to verify that if the other outcomes
of Alice's measurement are the rest of a complete basis of maximally
entangled states (equivalently, as shown e.g. in  \cite{Werner}, if
the other outcomes correspond to projectors  
$T_V := (V \otimes_{AB} I) T (V^\dagger \otimes_{AB} T)$ where $V$ varies over the
elements (except $I$) of an orthogonal unitary
basis for the local operators), then a similar result obtains:

\begin{theorem} \label{blah}
Let $T$ and $U$ be as described above. Then for 
any operator $W$ on $\H_1$,  
\begin{boxx}
\begin{array}{c}
(T_V \otimes_{AB} \1)(W \otimes_{12} T)(T_V \otimes_{AB} \1) = \alpha T \otimes_{AB} V^t W V^*.\\
\text{and}\\
(\1 \otimes_{AB} T_V)(W \otimes_{12} T)(\1 \otimes_{AB} T_V) = \alpha V W V^\dagger
\otimes_{AB} T.\end{array}
\end{boxx}
\end{theorem}

(Note that $V^*$ here means the operator whose matrix elements
(relative to some fixed basis which is also that used to define $V^t$)
are the complex conjugates of $V$'s, so that for any operator $V$, of
course $V^* \equiv (V^t)^\dagger$.)

The Theorem has the following immediate consequence:

\begin{corollary} \label{whatever}
Let $T$, $W$ and $U$ be as above (with $W$ normalized).  Suppose the
state $X = W \otimes_{AB} T$ is POPT on $\H_{A} \otimes \H_{B}$,
i.e., suppose that $\Tr((A \otimes_{AB} B)X) \geq 0$ for all positive
operators $A, B$ on $\H$. Then $W$ is positive.
\end{corollary}
 
 {\bf Proof:} Let $A = T$. Note that 
$T \otimes_{AB} B = (\1 \otimes_{AB} B)(T \otimes_{AB} \1)$. 
Thus, 
\begin{eqnarray*}
\Tr((T \otimes_{AB} B)X) & = & \Tr((T \otimes_{AB} B)X(T \otimes_{AB} B)) \\
& = & \Tr((\1 \otimes_{AB} B)(T \otimes_{AB} \1)X (T \otimes_{AB} \1)(\1 \otimes_{AB} B))\\ 
& = & \Tr((\1 \otimes_{AB} B)(T \otimes_{AB} W)(\1 \otimes_{AB} B)) \\
 & = & \Tr(WB).\end{eqnarray*}
 By assumption, this is non-negative for all positive operators $B$ on
$\H$; hence, $W \geq 0$.  (The third equality in the displayed
equations used Theorem \ref{bleah}, with $\alpha =1$ because $\tr W =
1$.) $\QED$

Thus, notwithstanding that $W$ and $T$ are POPT on $\H_1$ and
$\H_2$ respectively, if $W$ is not positive, then $W \otimes_{12}
T$ is not POPT on $\H_{A}$ and $\H_{B}$. This reflects the
fact that pure tensors in $\H_{A} \otimes \H_{B}$ typically
involve entanglements between $\H_{A_1}$ and $\H_{B_1}$.

In other words, if {\em locally}, i.e. at Alice's and Bob's sites,
systems combine according to the usual quantum rules, and in particular,
measurements with ``entangled outcomes'' like the outcome $T$ are permitted,
then POPT but non-positive states {\em cannot} occur as independent
states of ``nonlocal subsystems'' (subsystems such as $\H_{1}$, 
$\H_{2}$) of the Alice-Bob system.  Since specifying a Hermitian
operator such as 
$W \otimes_{12} T$ specifies all probabilities for outcomes
$\ket{w}\ket{x}\ket{y}\ket{z}$, and since the projectors onto these
span the Hermitian operators,  we have 
established Proposition \ref{ivory}.

It is worth noting that although Propositions \ref{bleah}
and \ref{blah} describe operators that could be interpreted
as the unnormalized {\em overall} conditional state after
Alice gets various Bell-measurement results ($T$ or $T_V$),  
if the ``projection-postulate'' dynamics applies, this does
not mean that one might get around the difficulty pointed out in
Corollary \ref{whatever} by supposing the actual conditional dynamics
are not described by the projection postulate.   The Propositions
are used in the Corollary only for the purpose of calculating 
probabilities of certain one-shot outcomes (that exhibit no Alice-Bob 
entanglement), and for these the evolution of the {\em probability}
state {\em after} the measurement, i.e. the probabilities of 
{\em subsequent} measurements, are irrelevant.  The ``projected''
states occur only under the trace in the calculations leading
to Corollary 3, and so are only used there in calculating the probabilities
of outcomes corresponding to the projectors, a proper application 
{\em whatever} the subsequent dynamics may be.

It is also worth noting that if only ``1-LOCC''
measurements and operations are permitted locally (where the
``locality'' of this 1-LOCC is now a ``fictitious'' locality with respect to
the product $\H_{A_1} \otimes \H_{A_2}$, or the product 
$\H_{B_2} \otimes \H_{B_1}$), 
then Alice cannot perform the 
measurement whose outcomes are maximally entangled states of 
$\H_{A_1} \otimes \H_{A_2}$, so the above argument causes no problem.
We conjecture (and it is probably simple to prove) that in this case 
the state $U (W \otimes T) U$ {\em is} a legitimate state, 
that is, a state on the test space $(XY, {\frak A}{\frak  B})$ of 
Definition \ref{def: influence-free product}, with $(X, \frak{A})$
and $(Y, \frak{B})$ themselves ``1-LOCC'' test spaces
$(X_1 X_2, \frak{A_1 A_2})$, $(Y_2 Y_1, \frak{B_2 B_1})$ made by 
combining standard quantum E-test spaces like $(X_1, \frak{A})$.

\textitem{Remarks:  further discussion of Corollary~\ref{whatever}}

One can envision a variety of reactions to this, including:
 \begin{itemize} 
\item[(A)] The argument shows that POPT but non-positive ``states" are
un-physical. 

  (In fact, if we apply the states/maps isomorphism,
we have here a non-standard proof of the
standard observation that the extension of a positive but not CP
map $\phi : {\cal L}(\H) \rightarrow {\cal L}(\H)$ to a map
${\cal L}(\H \otimes \H) \rightarrow {\cal L}(\H \otimes \H)$ ($\H$
finite dimensional) needn't be positive.)  
\item[(B)] The result shows
that non-positive POPT states on subsystems don't extend to POPT
states on larger systems. So what? Who says states on subsystems
should generally extend to states on larger systems?\\
\end{itemize}

As regards potential applications of the nonstandard tensor product
in physics, (A) seems relevant.  Of course, at least for most
physical situations we know quantum mechanics does work, and while we
have a reasonable amount of evidence for the existence of standard
quantum entangled states, we have so far observed no statistics like
those of the nonpositive but POPT states.  So even without bringing
in the mathematical considerations of (A), we might be disinclined
to consider such states.  However, the lack of evidence 
could, of course, just be ascribed to the fact that
we haven't looked hard enough, 
or that such 
states, for some reason, occur in exotic contexts not yet adequately
physically probed.  However, one then must take the attitude expressed
in observation (B)---but it may be difficult to reconcile this attitude
with existing
physics, including the existence of standard quantum entangled states
of subsystems seemingly independent from other systems. 

\textitem{Resum\'e}
The above results are inspired by an intuitive
argument that, within the FNS coupling of
A and B, teleportation gives rise to a post-Bell-measurement state
on Bob's side that would be nonpositive if Alice and Bob had teleported
a nonpositive POPT state.  In light of the results above,
we can make this more precise by saying that there 
there is no way to consistently do both of the following simultaneously:

\noindent
(1) assign probabilities to the outcome-pair
consisting of a Bell-outcome (such as would occur in the teleportation 
protocol) on Alice's side, and a particular ``locally 
entangled'' measurement outcome 
(dependent on the Bell outcome, although in teleportation this dependence
could be removed via classical communication and local adjustments made by Bob)
on Bob's side

\noindent
and 

\noindent
(2) assign to all outcomes that are not only pairs
of Alice and Bob outcomes, but in which Alice's outcome is 
a pair of an $A_1$ and an $A_2$ outcome (i.e. ``locally unentangled''), 
and similarly for Bob, the 
probabilities given by the product of
a Bell state of $A_1 B_1$, and a nonpositive 
POPT state of $A_2 B_2$.

It is nice that the argument does not 
depend on Bob's doing the required adjustments to actually get the nonpositive
POPT state on his side, for thus we can avoid issues of dynamics, of what
Bob can do in addition to measuring.  Rather, Bob's state conditional
on at least one (actually, all) of Alice's measurement outcomes is some
nonpositive 
POPT state, though only for one of the measurement outcomes is it the
POPT state that would be teleported in a full teleportation protocol
(otherwise it's an appropriate unitary transform of that state).

\section{Conclusion}

The results of our final section cast some doubt on the potential
relevance of nonpositive POPT states as models for undiscovered
physics, though they are hardly decisive
against this possibility.  Regardless of their relevance or lack of it 
as potential models for
phenomena in our physical world, though, nonpositive POPT 
states and the test spaces
they live on remain interesting and relevant for the theoretical
understanding of how systems can combine, and how this can affect
systems' information-processing capabilities\footnote{Other theoretical studies
along this line include Popescu, Rohrlich, and
collaborators work on nonlocal correlations \cite{Popescu-Rohrlich1994, 
Popescu-Rohrlich1996, Barrettetal} though there the restriction
to quantum marginals maintained in our work is absent in most cases.}.
Indeed, they even remain
relevant to our understanding of quantum mechanics, for they can be
interpreted as representing the failure of two particular very natural
way (and, it turned out, equivalent)
of trying to obtain the quantum-mechanical rules for constructing
composite systems: from local quantum mechanics and no-signalling, or
from local quantum mechanics and 1-LOCC.  This raises the question,
which we will discuss in future publications, of what additional natural
requirements (satisfied in the quantum mechanical cases) might be
imposed on notions such as test space or E-test space, so that
combining them and imposing no-signalling/1-LOCC 
gives the quantum tensor product.
Moreover, while this kind of system combination apparently cannot be
easily mixed with the standard quantum one (at least, not without
paying the price that product states do not universally exist,
endangering the interpretation of the factors as
``subsystems'' in the usual sense), it can still be used in a
thoroughoing way to combine systems in a fashion different from
quantum mechanics, and investigation of information processing in this
theory could illuminate general questions of what properties of a
theory are needed to do what information-processing tasks.

{\bf Acknowledgments}
We thank A. Khrennikov and the 
Center for Mathematical Modelling in Physics, Engineering, and
Cognitive Sciences at the University 
of V\"axjo for invitations to the conference ``Quantum Theory:
Reconsideration of Foundations, II,'' where some of the 
work was done, as well as the US DOE for financial support through
Los Alamos National Laboratory's Laboratory-Directed Research and 
Development (LDRD) program.


\begin{thebibliography}{} 
\bibitem{Barrettetal} Barrett, J., N. Linden, S. Massar, S. Pironio, 
S. Popescu and David Roberts, {\sl Nonlocal correlations as an 
information-theoretic resource}, Phys. Rev. A {\bf 71}, 022101, 2005.
\bibitem{Bunce-Maitland-Wright} L. J. Bunce and J. D. Maitland-Wright,
{\em The Mackey-Gleason Problem}, Bulletin of the AMS {\bf 26} (1992)
288-293.
\bibitem{Busch03a} P. Busch, {\em Quantum states and generalized
observables: a simple proof of Gleason's theorem},
Phys. Rev. Lett. {\bf 91}, 120403 (2003).
\bibitem{Busch-Lahti} Busch, P., and Lahti., P., {\em Remarks on
separability of compound quantum systems and time reversal},
Found. Phys. Lett. {\bf 10} (1997), pp. 113-117.
\bibitem{CFRM03a} Caves, C.~M. and C.~A. Fuchs and K. Manne and
J.~M. Renes, {\em Gleason-type derivations of the quantum probability
rule for generalized measurements}, Found. Phys. {\bf 34}, p. 193,
2004.
\bibitem{Choi} Choi, M-D. {\em Completely Positive Maps on Complex
Matrices}, Lin. Alg. Apl. {\bf 10} 285-290, 1975.
\bibitem{Dvurecenskij95} Dvure{$\check{c}$}enskij, A. {\sl Tensor product of difference
posets}, Trans. Am. Math. Soc. {\bf 346}, pp. 1043--1057.
\bibitem{Foulis-Bennet} D.~J. Foulis and M.K.~Bennet, {\sl Tensor products of
orthoalgebras}, Order {\bf 10} (1993) 271-282.
\bibitem{Foulis-Bennet94} D.~J. Foulis and M.K.~Bennet, 1994. {\sl Effect algebras and unsharp
quantum logics}, Found. Phys. {\bf 24}, pp. 1325--1346.
\bibitem{Foulis-Randall80a} D.~J. Foulis and C.~H. Randall, 
{\em What are quantum
logics and what ought they to be?}, in Beltrametti and van Fraassen
(eds.), {\bf Current Issues in Quantum Logic}, pp. 35--52.  Plenum,
1980.
\bibitem{Foulis-Randall80b} D.~J. Foulis and C.~H. Randall, 
{\em Empirical logic
and tensor products}, in B. Hartkamper and H. Neumann (eds.), {\bf
Interpretations and Foundations of Quantum Mechanics: Proceedings of a 
Conference hold {{\rm [sic.]}} in Marburg 29-30 May, 1979}, 
pp. 9--20.  Z\"urich: Bibliographisches Institut Wissenschaftsverlag, 1981.
\bibitem{Fuchs2001a} Fuchs, C.~A., {\em Quantum Foundations in the
light of quantum information}, in A. Gonis and P.~E.~A. Turchi (eds.),
{\bf Proceedings of the NATO Advanced Research Workshop on Decoherence
and its Implications in Quantum Computation and Information
Transfer.} IOS Pr Inc., 2001.
\bibitem{Fuchs2002a} Fuchs, C.~A., {\em Quantum mechanics as quantum
information (and only a little more)}, quant-ph/0205039.
\bibitem{FJ01a} C.~A. Fuchs and K. Jacobs, {\em Information tradeoff
relations for finite-strength quantum measurements}, Phys. Rev. A {\bf
63}, 062305 (2001).
\bibitem{Gleason} Gleason, A.~M., {\em Measures on the Closed Subspaces of
Hilbert Space}, J. Mathematics and Mechanics {\bf 6}, pp. 885--893 (1957).
\bibitem{Gudder97} Gudder, S., 1997. {\sl Effect Test Spaces}, {Int. J. Theoretical 
Physics}, {\bf 36}, p. 2681.
\bibitem{Horn-Johnson} Horn, R.~A. and Ch.~R. Johnson,
{\bf Matrix Analysis},
Cambridge University Press, 1985.
\bibitem{Horodeckis} M. Horodecki, P. Horodecki and R. Horodecki, 
Physics Letters A {\bf 223}, pp. 1--8, 1996. 
\bibitem{Jamiolkowski} A. Jamiolkowski, {\em Linear transformations
which preserve trace and positive semidefiniteness of operators},
Rev. Mod. Phys.  {\bf 3}, pp. 275--278, 1978.
\bibitem{Kadison-Ringrose} Kadison, R.~V. and J.~R. Ringrose {\bf
Fundamentals of the Theory of Operator Algebras}, vol. 1, 
{\sl Graduate Studies in Mathematics vol. 15}, Providence, RI: 
Americam Mathematical Society (1997). 
\bibitem{Klay} Kl\"{a}y, M. {\sl Einstein-Podolski-Rosen experiments:
the structure of the probability space I}, Found. Phys. Lett. {\bf 1},
pp. 205--244, 1988; {\sl Einstein-Podolski-Rosen experiments: the
structure of the probability space II}, Found. Phys. Lett. {\bf 1},
pp. 305--319, 1988.
\bibitem{Klay-Randall-Foulis} { Kl\"{a}y, M., Randall, C., and Foulis,
D., {\sl Tensor products and probability weights},
Int. J. Theor. Phys. {\bf 26} (1987) 199-219
\bibitem{Kopka} K{$\check{o}$}pka, F.  and F. Chovanec, 1994.  {\sl D-posets}, 
Mathematica Slovaca, {\bf 44}, pp. 21--34. 
\bibitem{Hellwig69a} K.~E. Hellwig and K. Kraus, {\em Pure operations
and measurements}, Commun. Math. Phys.}, {\bf 11}, 214-220 (1969).
\bibitem{Hellwig70a} K.~E. Hellwig and K. Kraus, {\em Operations and
measurements: {II}}, {Commun. Math. Phys.}  {\bf 16} 142-147 (1970).
\bibitem{Kraus} Kraus, {\bf States, Effects and Operations}, Springer
Lecture Notes in Physics {\bf 190}, Springer Verlag: Berlin, 1983.
\bibitem{Namioka-Phelps} Namioka, I., and Phelps, R. R. {\em Tensor
Products of Compact Convex Sets}, Pacific J. Math {\bf 9}, p. 469, 1969.
\bibitem{Pitowsky} I. Pitowsky, {\bf Quantum Probability/Quantum
Logic}, Springer-Verlag, 1989.
\bibitem{Popescu-Rohrlich1996} S. Popescu and D. Rohrlich, {\em
Nonlocality as an axiom of quantum theory}, Annals of the Israel
Physical Society, {\bf 12}, pp. 152--156, 1996.  (Also {\tt quant-ph/9508009}, 1995).
\bibitem{Popescu-Rohrlich1994} {\sl Quantum nonlocality as an axiom},
Foundations of Physics {\bf 24}, pp. 379--385, 1994.
\bibitem{Randall-Foulis} Randall and Foulis, {\em Operational
statistics and tensor products}, in H. Neumann (ed.), {\bf
Interpretations and Foundations of Quantum Theory} pp. 21--28, 
B.I. Wissenschaft:
Mannheim (1981).
\bibitem{Rudolf-Wright} O. Rudolph and J.D. Maitland Wright, {\em On
unentangled Gleason theorems for quantum information theory},
quant-ph/00004036
\bibitem{Stinespring} Stinespring, {\em Positive Functions on
$C^{\ast}$-algebras}, Proc. Amer. Math. Soc. {\bf 6} 211-216, 1955
\bibitem{Stormer} St\"{o}rmer, {\em Positive Linear Maps of Operator
Algebras}, Acta Math.{\bf 110} 233-278 1963.
\bibitem{Wallach} Wallach, N. R., {\em An Unentangled Gleason's
Theorem}, quant-ph/0002058 
\bibitem{Werner}, R. F. Werner, ``All
teleportation and dense coding schemes,'' J. Phys. A: Math. Gen. {\bf
34}, p. 7081, 2001.
\bibitem{Wilce90} Wilce, A., {\em Tensor Products of Frame Manuals},
Int. J. Theor. Phys. {\bf 29} (1990) 805-814.
\bibitem{Wilce92} Wilce, A., {\em The Tensor Product in Generalized
Measure Theory}, Int. J. Theor. Phys. {\bf 31} (1992) 1915-1928.
\bibitem{Wilce93} Wilce, A., {\em Spaces of Vector-Valued Weights on
Test Spaces}, ancient preprint.
\bibitem{Wittstock} Wittstock, G., {\em Tensor Products of Ordered
Vector Spaces}, in Hartk\"{a}mper and Neumann, {\bf Foundations of
Quantum-Mechanics and Ordered Linear Spaces}, Springer Lecture Notes
in Physics {\bf 29}, 1974.
\end{thebibliography}
\end{document}